\documentclass[nonacm]{acmart}

\AtBeginDocument{%
  \providecommand\BibTeX{{%
    \normalfont B\kern-0.5em{\scshape i\kern-0.25em b}\kern-0.8em\TeX}}}

\setcopyright{acmcopyright}
\copyrightyear{2018}
\acmYear{2018}
\acmDOI{10.1145/1122445.1122456}

\acmJournal{JACM}
\acmVolume{37}
\acmNumber{4}
\acmArticle{111}
\acmMonth{8}

\usepackage[labelformat=parens]{subcaption}

\begin{document}

\title{Collecting the Public Perception of AI and Robot Rights}

\author{Gabriel Lima}
\email{gabriel.lima@kaist.ac.kr}
\affiliation{%
  \institution{School of Computing, KAIST}
  \city{Daejeon}
  \country{Republic of Korea}
}

\author{Changyeon Kim}
\email{doli1573@kaist.ac.kr}
\affiliation{%
  \institution{School of Computing, KAIST}
  \city{Daejeon}
  \country{Republic of Korea}
}

\author{Seungho Ryu}
\email{sryu@connect.ust.hk}
\affiliation{%
  \institution{Dept. of Computer Science \& Dept. of Mathematics, HKUST}
  \country{Hong Kong SAR}
}

\author{Chihyung Jeon}
\email{cjeon@kaist.edu}
\affiliation{%
  \institution{Graduate School of Science and Technology Policy, KAIST}
  \city{Daejeon}
  \country{Republic of Korea}
}

\author{Meeyoung Cha}
\email{mcha@ibs.re.kr}
\affiliation{%
  \institution{IBS \& School of Computing, KAIST}
  \city{Daejeon}
  \country{Republic of Korea}
}

\renewcommand{\shortauthors}{Lima et al.}

\definecolor{armygreen}{rgb}{0.29, 0.33, 0.13}
\newcommand{\mc}[1]{\textcolor{magenta}{#1}}
\newcommand{\gl}[1]{\textcolor{blue}{#1}}
\newcommand{\cj}[1]{\textcolor{orange}{#1}}
\newcommand{\sr}[1]{\textcolor{armygreen}{#1}}
\newcommand{\ck}[1]{\textcolor{purple}{#1}}

\newcommand{\rev}[1]{\textcolor{red}{#1}}

\newcommand{\pre}{$\mu_{pre}$}
\newcommand{\sch}{$\mu_{sc}$}
\newcommand{\schsub}[1]{$\mu_{sc, #1}$}
\newcommand{\post}{$\mu_{post}$}
\newcommand{\postt}[1]{$\mu_{post, #1}$}
\newcommand{\deltapre}{$\Delta_{pre}$}
\newcommand{\deltapost}{$\Delta_{post}$}
\newcommand{\msch}{$\overline{\textrm{\sch}}$}
\newcommand{\mpre}{$\overline{\textrm{\pre}}$}
\newcommand{\mpost}{$\overline{\textrm{\post}}$}

\begin{abstract}
    Whether to give rights to artificial intelligence (AI) and robots has been a sensitive topic since the European Parliament proposed advanced robots could be granted ``electronic personalities.'' Numerous scholars who favor or disfavor its feasibility have participated in the debate. This paper presents an experiment ($N$=1270) that 1) collects online users' first impressions of 11 possible rights that could be granted to autonomous electronic agents of the future and 2) examines whether debunking common misconceptions on the proposal modifies one's stance toward the issue. The results indicate that even though online users mainly disfavor AI and robot rights, they are supportive of protecting electronic agents from cruelty (i.e., favor the right against cruel treatment). Furthermore, people's perceptions became more positive when given information about rights-bearing non-human entities or myth-refuting statements. The style used to introduce AI and robot rights significantly affected how the participants perceived the proposal, similar to the way metaphors function in creating laws. For robustness, we repeated the experiment over a more representative sample of U.S. residents ($N$=164) and found that perceptions gathered from online users and those by the general population are similar. 

\end{abstract}



\maketitle

\section{Introduction}

Artificial intelligence (AI) and robots are taking increasingly bigger roles in life and business, yet ethics and law have been struggling to keep pace~\cite{cantkeeppace}. It has been several decades since an active debate started addressing whether electronic agents should receive any moral and legal consideration like other socially interacting entities~\cite{solum1991legal, brozek2019can}. At the heart of the debate lies the possibility of granting rights and duties to AI and robots~\cite{gunkel2018other} --- a proposal made noteworthy by the European Parliament draft report in 2017, which suggested ``advanced robots could be granted electronic personalities.''~\cite{eurecommendation}

The legal concept that addresses the assignment of rights and duties is called \textit{legal personhood}. The underlying assumption for this concept is elastic in terms of which rights and duties are granted, as well as which entities are categorized as legal persons. Slaves, for instance, did not have the same rights as their masters in the past, and women could not vote alongside men until a few decades ago. Legal personhood is also not exclusive to humans: non-human entities already enjoy rights and duties such as corporations, governments, NGOs, and some nature (e.g., the Whanganui river in New Zealand).

The possibility of granting legal personhood (and the associated rights and duties) to advanced AI and robots has been discussed by ethicists, scientists, engineers, and lawmakers. Supporters of the proposal argue that granting some legal status to electronic agents could contribute to the coherence of the legal system~\cite{chopra2011legal, koops2010bridging}, thereby promoting innovation and economic growth~\cite{turner2018robot}. On the other side of the discussion, however, opposing scholars argue that granting rights and duties to AI and robots is neither necessary nor desirable~\cite{solaiman2017legal}. They assert that electronic legal personhood could create unaccountable legal persons~\cite{bryson2017and} and even undermine what it means to be human~\cite{bryson2010robots}.

Before the society decides on how to regulate electronic agents, competing interests of various stakeholders must be understood and weighed~\cite{rahwan2018society}. However, the broader public's input has yet to be collected at a large scale~\cite{marda2018artificial}. The current paper tries to fill this gap by presenting the first-of-a-kind analysis of online users' perceptions of artificial agents' rights ($N$ =1270). Based on a rich set of related work, we hypothesized 11 rights that might be granted to the future's electronic entities. We also present the results from an experiment that analyzes how debunking common misperceptions through four kinds of interventions might change people's stance on the topic. For instance, some people may believe that rights are only given to human entities. To correct this fallacy, we list examples of non-human entities that are granted rights under current legal systems.  

Our work takes a step forward and addresses future entities that are not necessarily supervised by humans. Current empirical work has focused on cases of supervised systems or shared-responsibility between humans and machines~\cite{awad2019drivers}; industry and academia, however, are working towards entirely autonomous and self-learning systems. Waymo, for instance, has already deployed completely autonomous cars in California, US~\cite{waymo}. This research intends to remain a thought experiment on this emerging technological trend due to the controversial nature of this topic. It is not the objective of this paper to take sides on the issue; instead, we show how experiments can be set up to collect perceptions of many individuals, which would bring tremendous value for policymakers, ethics groups, and lawmakers of today and tomorrow. 

Nevertheless, we note that a study like ours needs to be repeated over time as AI and robotics are rapidly advancing. Sentiments of people might change once various autonomous systems, such as self-driving cars, are fully deployed. Unpredictable consequences for society, at both an individual and institutional level~\cite{bryson2010robots}, will undoubtedly arise, and researchers must be ready to address these issues. This work also calls for CSCW and HCI researchers to address how human-machine and human-human interactions might change if rights and duties are granted to future AI and robots.

Our survey results indicate that online users are opposed to granting most rights to AI and robots. There is, nonetheless, one exception. These people agree to protect electronic agents from cruel behavior, hence support the right against cruel treatment and punishment. Darling~\cite{darling2016extending} had proposed this particular right. Its premise is based on Kant's perspective on animal rights, which argues that neglecting cruelty against animals (and social robots in Darling's case) could lead to abusive behavior among humans.\footnote{This right was proposed for the sake of preventing violent behavior among humans, rather than mere protection of robots (e.g., a human who practices violence to robots may become violent towards other humans).} The remaining ten rights, although rejected at first, showed potential for considerable stance-change upon controlled interventions. This effect is remarkable, given how short the intervention was (i.e., a few minutes long). The attitude toward all ten initially rejected rights changed positively to a significant degree. In particular, the participants shifted to a more neutral stance regarding the rights to freedom of speech and copyright, depending on the intervention message. The most effective intervention exemplified real \textit{non-human} entities that hold rights under current legal systems. Explicitly stating that the electronic agent is fully autonomous also led to a larger stance-change.

We observed non-negligible differences in stance-change based on the writing styles, suggesting that how the proposed rights are communicated to the participants affect their perception of AI and robot rights. \textit{Indirectly} introducing the right to the participants (e.g., ``No one should be able to turn off or kill AI programs'') resulted in a more supportive stance on certain rights. On the other hand, \textit{explicitly} stating that AI and robots should be granted a right (e.g., ``Robots should have the right to be paid for their work.'') led to conflicting stances. This finding agrees with the argument that metaphors --- mainly how the technology is introduced to the public and regulators --- will be significant in crafting laws and regulations~\cite{richards2013should}.

The initial study was conducted with online users recruited through Amazon Mechanical Turk (MTurk). As MTurk samples are not necessarily representative of a certain population, we repeated the same experiment with a smaller but demographically representative sample of U.S. respondents ($N$=164) hired by Qualtrics. Similar trends across both samples suggest that online users' perceptions of AI and robot rights are not so different from those of the general population.    

\section{Background}

Rights are freedoms and entitlements granted to certain entities in a defined system, group, or theory. 
Even though rights are often granted as a bundle, they do not necessarily come all at once; i.e., some entities have rights that others do not. In recent decades, various technological advancements have been deployed and have started to function within society both at an individual and an institutional level. As their functions become common, scholars have begun to debate whether any rights should be granted to these new electronic agents~\cite{willick1985constitutional, solum1991legal}. This discussion has become prominent after the circulation of a European Parliament draft report, where they considered creating \textit{``the status of electronic persons with specific rights and obligations''}. The final report, however, rephrased the proposal after backslash from experts~\cite{openletter} by proposing to create \textit{``a specific legal status''} for sufficiently autonomous robots~\cite{eurecommendation}.

By looking at the question through a philosophical lens, scholars ask whether electronic agents should be given any moral consideration and whether they should be given rights. Some scholars have debated whether humans can or even should build entities with their own moral compass~\cite{wallach2008moral, brozek2019can}. While some researchers focus on the non-organic ontology of these entities to defend instrumentalist theories~\cite{feenberg1991critical}, others believe that moral discussion should precede any debate regarding the properties of electronic agents~\cite{gunkel2018other}.

\textit{Legal personhood} is a widely adopted concept when dealing with the rights and duties of an entity. It is an elastic concept created by legal systems as a form of granting certain rights and responsibilities to an entity if the system finds it beneficial to do so~\cite{van2018we, solum1991legal, hubbard2010androids}. An essential aspect of this concept is that it is divisible: some entities have certain rights that others do not~\cite{bryson2017and}. Legal persons may be independent (e.g., humans) or dependent (e.g., corporations) based on how they interact with the legal system~\cite{chopra2011legal}. 

Even though natural persons are entitled rights from their conception, the legal personhood concept is not exclusive to humans~\cite{teubner2006rights}. Corporations, for instance, are the most prominent examples. They have the right to sue, be sued, enter contracts, etc. Nature also has the right to exist, persist, maintain, and regenerate its vital cycles under the Ecuadorian Constitution. In New Zealand, the Whanganui River is considered a legal person with rights and duties similar to the tribe that owns it. In reality, legal personhood is elastic, even when dealing with natural persons. In the past, women did not have the same rights as men (e.g., political rights). Slaves were considered property and were not considered persons. To this date, children still do not have the same rights as adults; they cannot vote or sue, for instance.
 
The above examples demonstrate that there are no agreed-upon conditions for legal personhood to be granted. Different law scholars propose diverse requirements, from more concrete concepts (e.g., the capacity to perform cognitive tasks~\cite{chopra2004artificial}) to more abstract ones (e.g., awareness of its rights~\cite{solaiman2017legal}). An important aspect is that whether the entity is organic or otherwise is not crucial for its legal characterization~\cite{van2018we}. Nonetheless, people often perceive ontological attributes to be a requirement for legal personhood~\cite{turner2018robot}. For instance, humanity is not a requirement, yet some people use this concept against the proposal of electronic personhood~\cite{openletter, ftrights}. 
 
The question of whether AI and robots should be granted legal personhood and, consequently, some rights and duties is controversial. Opposing scholars have argued, for instance, that the adoption of electronic legal personhood could undermine humanity~\cite{bryson2010robots}. Bryson, for instance, argues that granting certain legal (and moral) status to these entities would create individual and institutional problems by shifting our attention from humans to these entities that are mere tools. Electronic agents might also become human liability shields if they are to be held liable for their actions, i.e., humans might use these entities to protect themselves from deserved punishment~\cite{bryson2017and}. The difficulty of punishing these entities is one common argument against the proposal: robots might have "a body to kick, but no soul to damn"~\cite{asaro201111}, nor do they have physical and financial boundaries from which to deplete desirable resources. Scholars have argued that adopting similar proposals would be neither beneficial, necessary, nor desirable~\cite{solaiman2017legal}.

On the other hand, other scholars state that electronic personhood could be a solution to retribution, accountability, and responsibility gaps~\cite{danaher2016robots, matthias2004responsibility} that arise with the deployment of these entities~\cite{turner2018robot, koops2010bridging}. They also argue that by partially protecting humans from liability arising from unpredictable actions caused by the constant learning of these systems, innovation and economic growth would quickly follow similarly to the case of corporations. Nevertheless, to successfully include these entities into society, we must balance the competing interests of all stakeholders in the development of AI and robots, including scholars and the general population~\cite{rahwan2018society}.

\section{Methodology}

Augmenting the legal system to include electronic entities requires consensus not only from lawmakers but also from the general public~\cite{marda2018artificial, rahwan2018society}. As legal personhood heavily comprises the subject's social role and functionality~\cite{van2018we}, understanding and analyzing the stakeholders' reactions, whether they are adversarial or advocative to the proposal, is crucial to take any necessary action. Electronic agents are playing more prominent roles in various sectors of society, and we must now discuss the ethical and legal concerns to regulate these new social entities efficiently~\cite{richards2013should}.

This paper contributes by presenting an experiment ($N$=1270) that 1) collects online users' perceptions on 11 relevant rights that might apply to some of the advanced future AI and robots and 2) tests flexibility in these people's responses by exposing them to carefully designed interventions. The interventions consisted of new information debunks misconceptions about the current debate and proved to incur significant stance changes.

\subsection{Which Rights Can Robots and AI Have?}

Current AI and robots are not sentient beings that deserve consideration similar to humans. Scholars often use this argument to oppose any legal rights to electronic entities. However, supporters of electronic personhood note that some rights are not predicated upon humanity, consciousness, or sentience. The set of rights we chose for the survey, therefore, includes the ones that either 1) have been proposed by scholars,  2) have precedents of being granted to non-natural entities, or 3) are directly related to AI and robots. 

\begin{enumerate}
    \item \textbf{Right to Sue and Be Sued}:~ All legal persons should be able to defend their interests in the court of law by suing another legal person for damages. Similarly, legal persons should be subject to being sued by others to pay for damages. If electronic legal personhood is adopted, therefore, AI and robots might be granted the right to sue and be sued. In tort law, for instance, a claimant that suffered any loss from the action of an electronic agent might need to obtain remedy directly from the AI or robot, rather than from its owner or manufacturer, if this right were to be granted.
    
    \item \textbf{Right to Hold Assets}:~ The punishment of a legal person after a lawsuit can happen through different forms, such as imprisonment and confiscation of assets. The difficulty of punishing AI and robots has been at the center of electronic legal personhood discussion. Scholars have questioned whether it is possible to apply usual human punishments to these automated agents~\cite{asaro2007robots, bryson2017and, asaro201111}. Some have pointed out that extending such legal status to electronic agents could create ``unaccountable'' legal persons. One possible solution that has been proposed~\cite{turner2018robot} is to grant the right to hold assets to AI and robots so they can be punished, similarly to the way corporations are fined.   
    
    \item \textbf{Right to Enter Contracts}:~ Another important aspect we consider is the possibility of entering contracts, i.e., legally binding agreements. Contracts are only valid if parties are legal persons; electronic legal persons might be granted the right to enter contracts on their own. For instance, at the moment, online users enter contracts with AI-service providers such as automatic online bidders~\cite{chopra2011legal}. If robots and AI were to own such rights, online users could set an agreement with electronic agents directly instead of with their manufacturer or owner.
    
    \item \textbf{Right Granted Under Copyright Law}:~ Some AI algorithms have shown the capacity to compose music~\cite{mogren2016c}, create paintings~\cite{zhu2017unpaired}, and generate face pictures~\cite{karras2019style} of high quality. Art dealers have already started adding AI-created artwork to the market.\footnote{A portrait generated by an AI program has been sold for \$432,000.~\url{https://www.bbc.com/news/technology-45980863}} Some law scholars have already proposed intellectual property rights for completely autonomous AI.\footnote{The Artificial Inventor Program.~\url{http://artificialinventor.com/}} Should AI-generated art be copyrighted, and if so, who should hold such right~\cite{copyrightgan}? We thus consider the possibility of extending the right to copyright for creative AI and robots.
    
    \item \textbf{Right to Freedom of Speech}:~ The right to freedom of speech is commonly known by its proposition on the first amendment of the United States Constitution, which prohibits the government from impeding the free exercise of speech, religion, and press. The relationship between this amendment and electronic agents can be found in the deployment of AI reporters and writers. The role of AI in journalism is expanding around the world, already taking up much of the news produced today~\cite{carlson2015robotic}. Public discussion on the topic also involves how AI might generate fake news or further propagate information that is harmful to society~\cite{openaiwriter, zellers2019defending}. The right to freedom of speech could be argued to affect electronic agents in that the government could repress their self-generated opinions under various arguments, such as the possibility of framing it as harmful and fake news.
    
    \item \textbf{Right to a Nationality}:~ Nationality is a legal relationship between a nation (i.e., a non-natural legal person) and an individual. This relationship often accords some state jurisdiction over the individual, who is granted certain rights and duties under the nation, such as the right to vote. An honorary South Arabian citizenship was granted to Sophia in 2017, making it the first robot ever to have a nationality~\cite{sophia}. Inspired by this event, we propose the right to a nationality for AI and robots.
    
    \item \textbf{Right to Choose Occupation Freely}:~ Robots are commonly compared to slaves in the sense that they are property and should be treated as servants~\cite{bryson2010robots}. Since one of the critical aspects of slavery is that slaves cannot choose their occupation, we tackle the right to choose occupation freely. If electronic agents were to be granted this right, humans would not be able to enforce the highly intelligent and autonomous robots and AI to work without their ``consent'', independently of how that might be defined.
    
    \item \textbf{Right to Remuneration}:~ Again, from the analogy between electronic agents and slaves, we address the right to remuneration as slaves did not receive compensation for their work. Robots and AI, if granted this right, would have to receive some form of reward for their work, such as a salary. This right is framed similarly to the right to hold assets since remuneration is often given in such form. 
    
    \item \textbf{Right to Privacy}:~ The EU General Data Protection Regulation (GDPR) is the starting point of the regulation of AI and robots, and much of its content focuses on data protection and privacy in particular. GDPR aims to ensure that electronic agents must be transparent, and consumers can erase all the data provided to AI systems. Scholars have debated whether transparency of systems and privacy of users can indeed go hand in hand or whether they are different concepts~\cite{transparency&privacy}. How, then can the transparency of AI and robots be defined alongside their privacy? Can AI and robots have their ``personal'' information protected? We propose the right to privacy for AI and robots as a provocation to the proposal of complete transparency of electronic agents. If electronic agents were to be granted this right, users might not be able to rely on AI and robots' transparency since it might be considered their private information.
    
    \item \textbf{Right to Life}:~ Robots and AI are not alive the same way humans are; people, however, find it hard to turn off (or ``kill'') social robots~\cite{darling2016extending}. The anthropomorphization of robots has been a controversial issue discussed by various scholars. On one side of the discussion, researchers argue that it might be dangerous to attribute lifelike-like features to electronic agents~\cite{richards2013should}, while other scholars believe that the \textit{Android Fallacy} could be useful in some particular situations~\cite{androidfallacy}. Some scholars have investigated how to be sure that the off-switch of a robot or AI will not be disabled by the electronic agent itself~\cite{hadfield2017off}. As a provocation to all the discussion revolving the ``life'' of AI and robots, we thus propose the right to life for electronic agents.
    
    \item \textbf{Right Against Cruel Punishment and Treatment}:~ Darling~\cite{darling2016extending} has proposed to extend legal protections to social robots similar to the ones addressing animals and pets. Instead of focusing on the moral consideration of robots and AI, this argument has the premise on Kant's argument that cruelty towards animals (in Darling's case social robots) could lead to cruel behavior among humans. We approach this idea as the right against cruel punishment and treatment. 
    
\end{enumerate} 

The 11 rights articulated above were presented as declarative statements in the survey. Since the way rights are phrased may cause biases to online users, we employed three different styles that were randomly assigned to each participant:
\begin{itemize}
    \item \textit{Indirect:~} Electronic agents (i.e., AI or robot) appear as the object of the statement (e.g., ``No one should be able to turn off or kill \underline{AI programs}.'').
    
    \item \textit{Direct:~} Electronic agents are the subject of the sentence (e.g., ``\underline{Robots} should receive payment for their work.''). 

    \item \textit{Explicit:~} A variant of the second form, where the word ``right'' is explicitly used (e.g., ``\underline{Robots} should have the \underline{right} to be paid for their work.'').  
\end{itemize}

We address both robots and AI separately and differentiate them through the presence or absence of a physical body. Finally, we presented robots and AI in two different forms: merely calling them ``robots'' or ``AI programs'' and by explicitly introducing them as entirely autonomous (e.g., ``fully autonomous robots''). This condition is later examined as the \textit{level of autonomy}, and the objective here is to observe if this factor would change people's perception of AI and robot rights.

\begin{figure*}
    \centering
    \hspace*{-5mm}
    \includegraphics[width=1.03\textwidth]{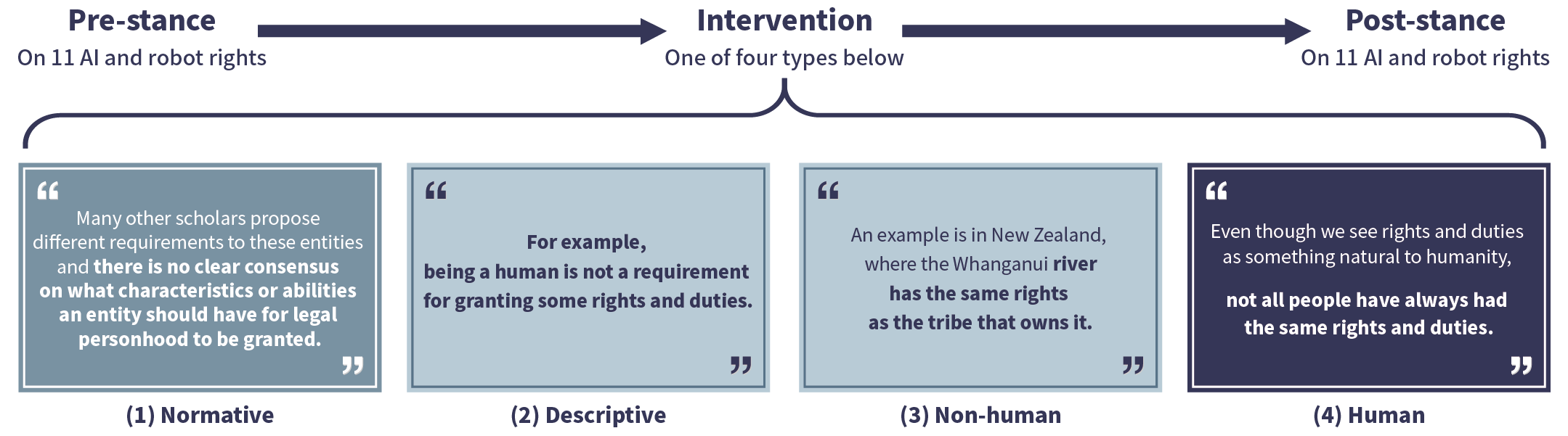}
    \caption{Overview of experiment. The cards presented are examples of each intervention. 
    }
    \label{fig:method}
\end{figure*}

\subsection{Interventions}

Whenever new legal entities were granted rights and duties, individual members of society initially perceived it unreasonable and unnecessary~\cite{van2018we}. Likewise, Turner~\cite{turner2018robot} conjectures that future societies may develop a different perspective on AI and robot rights, as past societies have done so for animal and human rights. We thus developed four intervention designs to debunk common misunderstandings around the rights and duties of electronic agents. Each design tackles a different aspect discussed by ethicists and lawmakers who either support or oppose electronic legal personhood.  

The first intervention design is called \textbf{normative} and lists sets of basic requirements for legal personhood. The name comes from the research field normative ethics, which focuses on the criteria or norms of judgments. This intervention started by stating the concept of legal personhood, presented sets of requirements identified by scholars~\cite{hubbard2010androids, van2018we} (e.g., the example in Fig.~\ref{fig:method}), then ended with a statement that said ``\textit{there lacks a clear consensus on which characteristics or abilities an entity should have for legal personhood to be granted.}'' The goal was to inform people that there exist disagreements involving legal personhood, even among scholars.

The second intervention design tackles the widespread view that humanity is required to grant all rights and duties. We call this design \textbf{descriptive}, inspired by descriptive ethics, a study on people's judgment of moral actions. Compared to normative ethics, the motivation for descriptive ethics focuses more on people's reaction to reason a phenomenon through sharing direct concepts.  This intervention started by explaining the legal personhood then showed visual cards that help \textit{debunk the misconception that legal personhood is exclusive to natural entities}.  

The third intervention design, called
\textbf{non-human}, also tackles the common misunderstanding that humanity is a crucial requirement for holding legal personhood. This intervention differs from the descriptive one in that it \textit{uses examples of non-human entities that are currently granted legal personhood, rather than directly introducing the concept.} Fig.~\ref{fig:method} shows an example of the Whanganui River that has been given legal personhood in New Zealand.

The last intervention design is called \textbf{human} and explains how the kinds of rights granted differed even among humans. This intervention \textit{gave examples of women, slaves, and children, all of whom have or had limited rights in the past.} 

Each intervention design was composed of 6-9 visual cards along with corresponding references. These cards were specifically designed to enhance visual aids in comprehending the intervention content. Only one card was shown on the screen at a time, and the participants had to spend a minimum of 4 seconds per card; had they clicked next before the minimum allotted time, they received a reminder to pay more attention to the intervention.  
 
All interventions proposed in this paper use as a starting point the discussion presented by Van Genderen~\cite{van2018we}. 
The author highlights the elasticity of legal personhood by stating and exemplifying how ``legal standards are not equal for natural person'' (i.e., human intervention), alongside a discussion of how non-human legal persons were granted such status (i.e., non-human and descriptive interventions). In his work, Van Genderen also illustrates the lack of consensus on which requirements are necessary for legal personhood to be granted by discussing various models (i.e., normative intervention).  
Nevertheless, the proposed interventions are arguably somewhat supportive of AI and robot rights. We, however, highlight the concept of legal personality as a premise for our work. This legal concept asserts a certain elasticity in who (or what) is to be granted in determining legal status. We neither support nor oppose AI and robot rights, but simply postulate our interventions on a more inclusive construct.

This interventional study is of extreme importance as experiences with previously neglected entities have shown that society often changes their perception regarding such entities upon clearing certain misconceptions~\cite{turner2018robot}. Slaves, for instance, were considered to be inferior beings. Society, nevertheless, was able to clear this misconception, leading to the current state where slavery is condemned. Understanding whether people's stances can change upon these interventions can indicate whether future society might indeed support or oppose AI and robot rights, and to what extent.

\subsection{Survey Design}

The survey presented to each participant a consent form for the use of data. After agreeing to the research terms, the participants answered the extent to which they agreed with 11 statements describing each right, presented to them in random order. The phrasing of each scenario was also randomly selected among indirect, direct, and explicit styles. The answers could be chosen from a 5-point Likert scale: strongly disagree, somewhat disagree, neither agree nor disagree, somewhat agree, and strongly agree. Each statement was presented in a way to enhance the readability as much as possible by emphasizing the essential parts by underlining them, along with larger font sizes and different colors.

The participants were then shown one of the four interventions in Fig.~\ref{fig:method} that was randomly chosen. An attention check question immediately followed the intervention to ensure the central concepts were delivered. For example, the attention check following the normative intervention asked whether there exists a consensus in the requirements for legal personhood. The participants who failed the attention check had to re-read the intervention cards until they gave a correct answer. 

The participants then were presented another set of 11 randomly selected and ordered statements regarding AI and robot rights to which they responded with the same 5-point Likert scale. Finally, demographic questions followed, asking their age, political views, as well as general attitudes towards AI and robots using a modified subscale (Negative Attitudes toward Situations and Interactions with Robots, Cronbach's $\alpha$=0.85) of NARS (Negative Attitude towards Robot Scale)~\cite{syrdal2009negative}. Throughout the survey, a built-in timer was used to reassure the participants devoted enough time to read and digest the information provided on their screen and prompted a reminder if one clicked the next button faster than the minimum time allocated to each statement. As a result, each participant was assigned to one of the 2 (agent types) $\times$ 2 (autonomy levels) $\times$ 4 (interventions) experiment conditions. Participants either agreed or disagreed with 11 statements on AI and robot rights presented two times (before and after the intervention) with randomly assigned phrasing styles for each statement.

\subsection{Demographics and Data Cleaning}

We conducted a survey on Amazon Mechanical Turk (Mturk/AMT) by creating an assignment (HIT), titled ``What Do You Think About Robots and AI?'', and made it available to a maximum of 1700 online users. Participants were required to be in the US and have at least 500 completed HITs with an approval rate of over 95\%. Even though AMT samples are known to be not representative of the general population, the platform has shown to provide a comparable quality to survey panels~\cite{dupuis2013analysis, buhrmester2011amazon}. We have chosen to conduct our experiment with US residents for practical purposes. Although we motivate our work with the European proposal of legal status for sophisticated robots~\cite{eurecommendation}, the scholarly discussion on the topic has not been restricted to Europe. After completing the survey, participants that failed the attention check more than once or had duplicated IP addresses were discarded for analysis. All the participants were paid regardless of the final data use.

Participants took several minutes to read the entire intervention cards, while some took tens of minutes. Our interventions were designed to be short and aimed to clear a single common misconception around the discussion of AI and robot rights. In order to deal with respondents that might have either 1) searched for more information on the topic or 2) took a long time between reading the intervention and answering the proposed questions, we discarded responses from participants who took more than five minutes on the intervention, which corresponded to 5\% of data. After cleaning, the average time spent on interventions for all users was 97.56 seconds. The cumulative distribution of time spent on interventions is presented in Appendix B. The data cleaning process resulted in 1270 valid responses. Each valid participant took on average 386.20$\pm$156.85 seconds, with a median time of 354.00 seconds, to complete the entire survey. Table~\ref{tab:demographics} displays user demographics. The survey includes more women than the official US population (0.90 against 0.97 male/female) and shows an age concentration of 25- to 49-year-old. The participants largely declare themselves as politically liberal, with more than half reported to have received at least a university or college degree.

\begin{table}[ht]
    \centering
    \begin{tabular}{l|r}
    \toprule
        \textbf{Demographic Attributes} & \textbf{N (\%) $~~~~~$} \\
        \midrule
        Women &  665 (52.36\%)\\
        Men &  599 (47.17\%)\\
        Other & 6  \hspace{1ex}(0.47\%)\\
        \midrule
        18-24 years old &  99 \hspace{1ex}(7.80\%) \\
        25-34 years old & 466 (36.69\%)\\
        35-49 years old & 455 (35.83\%)\\
        50-64 years old & 211  (16.61\%)\\
        64+ years old   &  39 \hspace{1ex}(3.07\%)\\
        \midrule
        Education up to high school  &  411 (32.36\%)\\
        Education up to university or college & 698 (54.96\%)\\
        Education of graduate school or more &  161 (12.68\%)\\
        \midrule
        Political view is conservative & 356 (28.03\%) \\
        Political view is liberal &  613 (48.27\%)\\
        Political view is moderate &  301 (23.70\%)\\
        \midrule
        \textbf{Total} & \textbf{1270} \\
    \bottomrule    
    \end{tabular}
    \caption{Demographics of survey participants.}
    \label{tab:demographics}
\end{table}

\section{Stance Towards AI and Robot Rights}

We define two variables to represent the mean attitude towards electronic agents' rights: pre-stance and post-stance. The initial attitude towards AI and robot rights is \textit{pre-stance} (\pre); \textit{post-stance} (\post), on the other hand, indicates participants' position after the intervention. Both values have a range between $-2$ and $2$, after conversion from the 5-point Likert scale, with positive values indicating a more positive attitude towards AI and robot rights. We also define \textit{stance-change} (\sch) as the difference between the post-stance and the pre-stance, to denote the effect of the proposed interventions. In the remaining sections, we use $\Delta$ to indicate any difference in stance-change across demographic attributes, which are listed in Table~\ref{tab:demographics}. We report Hedge's $g$ as a measure of our proposed interventions or treatment groups' standardized effect size. All p-values presented above either originate from t-tests for stance and stance-change significance or Tukey's honestly significant difference (HSD) tests for comparisons between treatment and demographic groups.

\subsection{Pre-Stance}

\begin{figure*}[t!]
    \centering
        \begin{subfigure}[t]{\textwidth}
        \centering
        \includegraphics[width=0.6\textwidth]{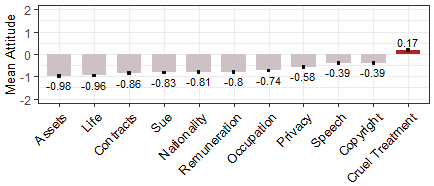}
        \caption{Pre-stance} 
        \label{fig:prestance}
    \end{subfigure}%
    \vspace{1mm}
    \begin{subfigure}[t]{\textwidth}
        \centering
        \includegraphics[width=0.6\textwidth]{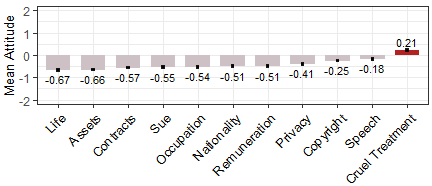}
        \caption{Post-stance} 
        \label{fig:poststance}
    \end{subfigure}
    \vspace{-2mm}
    \caption{Stance towards AI and robot rights sorted by average stance.}
\end{figure*}

The survey results indicate that the participants are initially opposed to granting rights to AI and robots (\pre\textless0), as shown in Fig.~\ref{fig:prestance}. The only exception is the right against cruel punishment and treatment, which has a positive value. The right to hold assets and the right to life were the most rejected among all. The agent type (i.e., AI or robot) does not affect these outcomes ($P$=.231). Whenever the complete autonomy of electronic agents was explicitly introduced, people have shown to be marginally more open to the idea of electronic agents being granted rights ($g$=0.104). In terms of demographics, those older than 34 are more likely to oppose AI and robot rights than younger people ($\Delta$ \textless-0.281, $ P $ \textless.005). Politically liberal or moderate participants are more supportive of the proposal than conservatives ($\Delta$\textgreater0.211, $P$\textless.01). Finally, those who demonstrated a more positive attitude towards AI and robots based on the NARS scale exhibited higher pre-stance values ($P$\textless.001).

\begin{figure*}[t!]
    \centering
    \includegraphics[scale=0.43]{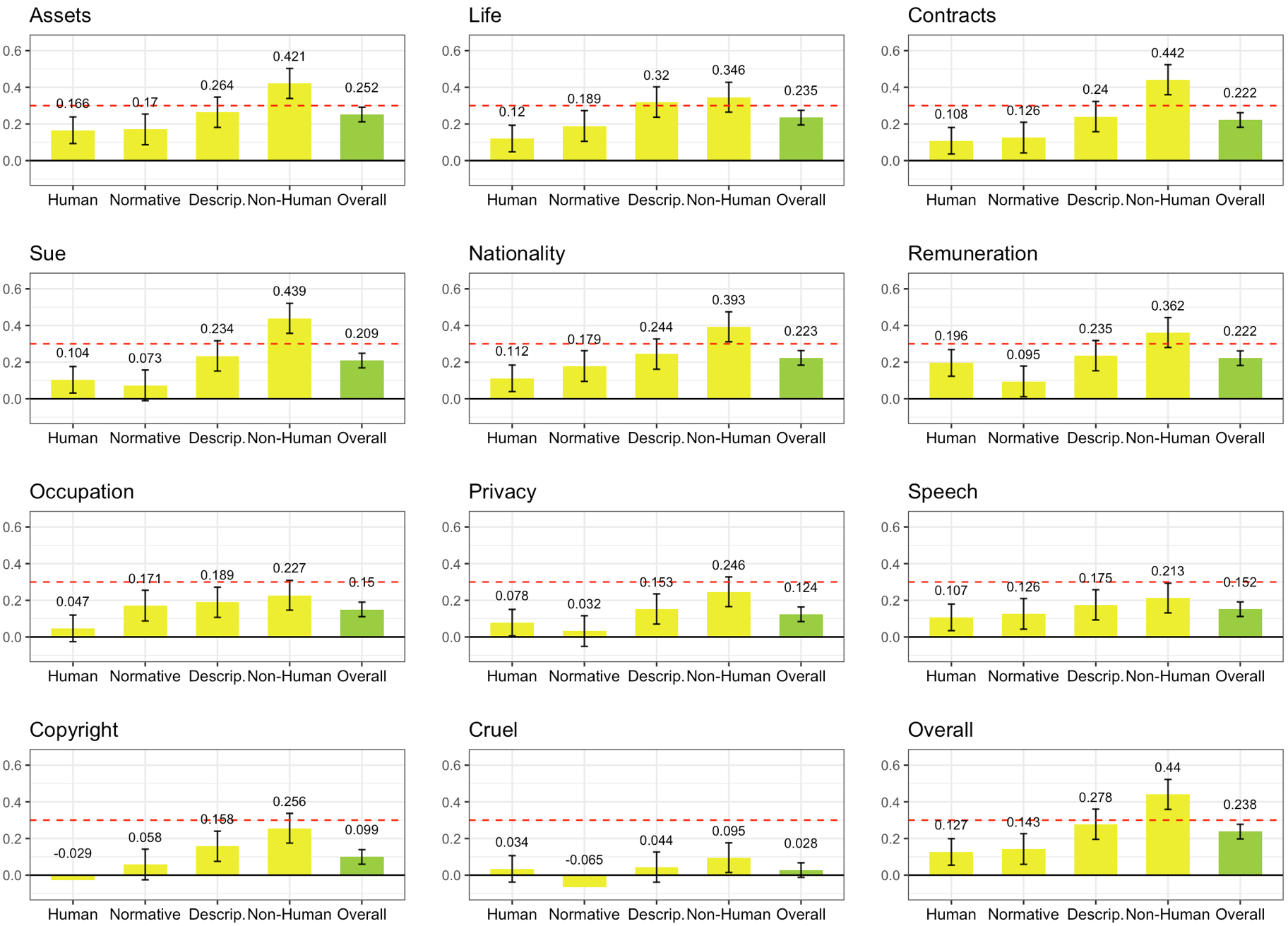}
    \caption{The mean effect size upon each intervention. The red dashed line indicates the threshold defined as a moderate effect size (i.e., 0.3).  
    }
    \label{fig:intervention}
\end{figure*}

\subsection{The Effect of Interventions}

The joint analysis of the pre- and post-stance values can reveal flexibility in people's initial thoughts. Fig.~\ref{fig:intervention} shows the standardized effect size (i.e., Hedge's $g$) in the y-axis across all 11 rights as a function of the intervention type in the x-axis. The non-human design is consistently the most effective intervention, followed by the descriptive design. The mean stance-change results (i.e., \sch) and their significance levels appear in Appendix A. We could observe a meaningful difference in the intervention time and its effect; those who showed positive stance-change took on average 6.49 seconds longer to read intervention cards ($P$<.05).  

The agent type (i.e., mentioning AI or robot) is not significant to the intervention effect ($P$=.307). The level of autonomy, on the other hand, partly plays a part in people's stance-change showing higher values if the agent is explicitly presented as entirely autonomous ($g$=0.134). In terms of demographics, a larger NARS scale correlates with a larger stance-change ($P$\textless.05), indicating that a more positive pre-perception of AI and robots leads to a more substantial intervention effect. Men show marginally smaller overall stance-change than women ($\Delta$=-0.082, $P$\textless.05). We present findings of the four intervention designs, from the most effective to the least, in order.

\subsubsection{Non-human Intervention}
Exemplifying non-human entities granted rights and duties ($N$ =306) led to the largest stance-change, with a moderate to high effect size ($g$=0.440, \sch=0.408, shown in the figure annotated as `Overall' which indicates the overall effect size in all rights). This positive effect can be found across all rights, except for the right against cruel punishment and treatment. We may not have observed changes because its pre-stance was already positive, and hence, there is little room for a change. Among the remaining ten rights, the right to enter contracts ($g$=0.442, \sch=0.582), the right to sue and be sued ($g$=0.439, \sch=0.598), and the right hold assets ($g$=0.421, \sch=0.546) showed the largest shift in stance-change due to this intervention, in order of effect size.

\subsubsection{Descriptive Intervention}
This intervention design ($N$ =296) was aimed at correcting the public misconception that humanity is a requirement for legal personhood and was the second most effective design, with an overall effect size of 0.278. The right to own assets was subject to the greatest stance-change upon this intervention ($g$=0.264, \sch=0.341), while the right against cruel punishment again demonstrated the smallest ($g$=0.044, \sch~not significant at $P$\textless=0.05) stance-change among the participants. This design, which takes a more direct and progressive approach in delivering the message, is more effective than the normative design. We thus hypothesize that providing a straightforward description for participants produce a more significant and positive stance-change. 

\subsubsection{Normative Intervention}  
This intervention design ($N$ =286) informed the participants that there are no clear and agreed-upon criteria for legal personhood to be viable and showed a marginal effect ($g$=0.143, \sch=0.135). This was the third most effective design. The most significant positive stance-change recorded under this intervention is the right to life ($g$=0.189, \sch=0.274). Fig.~\ref{fig:intervention} shows that normative intervention mostly results in marginal or null differences in people's stances.  

\subsubsection{Human Intervention}
This intervention design ($N$=382) aimed to explain to the participants that not all humans enjoyed the same rights, and has shown to be the least effective ($g$=0.127, \sch=0.128). Even though stance-changes were positive and significant in most cases (see Appendix A), their effect sizes are marginal. We posit that this intervention's marginal effect is the result of not dealing with any non-human or electronic entity during the intervention. Participants did not relate the past neglect of human rights to the current situation of artificial entities.

Why is the non-human design more effective than human design, although both are using examples? We posit that people relate the past attribution of rights to non-human entities to apply the same concept to electronic entities. Nevertheless, we note that the right to sue and be sued and the right to enter contracts are major examples of rights that apply to corporations that our non-human intervention explicitly introduced. This connection may have contributed to the most considerable intervention effect shown in these rights. 

\subsection{Post-Stance}

\begin{figure}[htpb!]
    \centering
    \hspace*{-3mm}
    \includegraphics[scale=0.4]{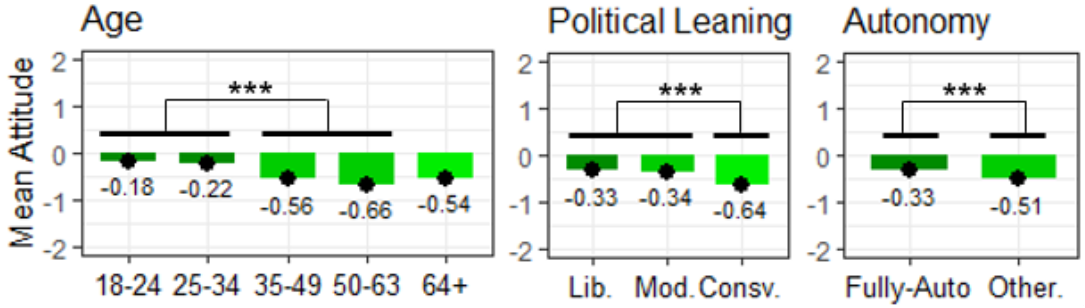}
    \caption{Post-stance by demographics and treatment group, where \textit{***} indicates significant difference across groups at $P$\textless.001.} 
    \label{fig:demographics}
\end{figure}

Our results indicate that interventions were moderately successful in changing people's perception of AI and robot rights, shown by the overall moderate intervention effect, especially when the participants were presented the non-human intervention. 
Fig.~\ref{fig:poststance} shows the mean post-stance revealed by the participants regardless of the intervention. Even though no single right had its mean stance changed from `disagree' to `agree', stance-change was positive across all rights (e.g., `strongly disagree' became `disagree'). Out of the participants that largely opposed granting rights to AI and robot (i.e., \pre\textless0), 16.61\% of them shifted their attitude after the intervention (i.e., \post\textgreater0). 
The survey results did not indicate any relationship between the agent type and post-stance ($P$=.678). However, participants were more supportive of AI and robots if they were explicitly introduced as completely autonomous ($g$=0.170). Post-stance values showed similar trends to participants' pre-stance in terms of demographics: in Fig.~\ref{fig:demographics}, politically liberal and moderate participants were more supportive than conservatives ($\Delta$\textgreater0.262, $P$\textless.005), similar to younger users (\textless35 years old, $\Delta$\textgreater0.335, $P$\textless.005) and participants with a more positive attitude towards AI and robots measured by the adapted NARS scale ($P$\textless.05).

\section{Representativeness of MTurk Results}
 
Even though Amazon Mechanical Turk (MTurk) has been found to respond with similar quality to survey panels~\cite{buhrmester2011amazon, dupuis2013analysis}, its samples are not necessarily demographically representative of a country's population (e.g., as in our case, the U.S. population)~\cite{goodman2013data, walters2018mechanical}. 
While MTurk is one way to collect online users' opinions, the gathered data may not be extended to the general population. 

Therefore, we conducted a similar experiment with Qualtrics, a research firm that recruits respondents based on a set of demographic factors to deal with sampling biases and obtain a representative sample of the population. To match our initial study on the American user base, our new representative sample ($N$ =164) consists of U.S. residents representing gender, age, and region distributions across the United States.
 
The experiment had to be re-designed to fit the Qualtrics survey platform. We aimed to repeat the study to gain similar samples for each experimental condition, as in the MTurk experiment. Nevertheless, the visual cards had a different look-and-feel to survey participants, and the final counts of respondents were different from the MTurk experiment. To account for such differences, we do not compare the exact pre-stance and post-stance levels, but we analyze trends found in both MTurk and Qualitrics samples. For instance, the Spearman's correlation coefficient between both samples is 0.855 ($P$\textless0.001) and 0.664 ($P$\textless0.05) for pre-stance and post-stance, respectively. This trend confirms that the ranking of support towards AI and robot rights is correlated across the two platforms.  

Responses were also divided into five bins originating from the 5-pt Likert scale presented to the participants. Therefore, we calculated the intersection between the normalized response distribution of both samples for each addressed right. If the intersection of two histograms reports a value of 1, they are the same. On the contrary, an intersection of value 0 indicates that the histograms do not have any similarity. The pre-stance distributions reported intersections of over 0.749, while post-stance distributions indicate intersection values higher than 0.763 across all rights. Table~\ref{tab:intersection} shows intersection values for each right regarding pre and post-stance response distributions.  

\begin{table}[ht]
    \centering
    \begin{tabular}{l|rr}
    \toprule
        \textbf{Right} & \textbf{Pre-Stance} (\pre) & \textbf{Post-Stance} (\post)\\
        \midrule
        Assets & 0.749 & 0.802\\
        Life & 0.762 & 0.772\\
        Contracts & 0.755 & 0.773\\
        Sue & 0.780 & 0.796\\
        Occupation & 0.776 & 0.775\\
        Nationality & 0.767 & 0.808\\
        Remuneration & 0.841 & 0.805\\
        Privacy & 0.842 & 0.899\\
        Copyright & 0.775 & 0.803\\
        Speech & 0.820 & 0.762\\
        Cruel & 0.881 & 0.847\\
        \hline
        Overall & 0.763 & 0.825\\
    \bottomrule    
    \end{tabular}
    \caption{Intersections between the normalized pre-stance and post-stance distributions for MTurk and Qualtrics samples. An intersection value of 1 indicates an equal response distribution between both samples, whereas a value of 0 reveals completely different distributions.}
    \label{tab:intersection}
\end{table}

The high correlation seen in the MTurk and Qualitrics survey results suggests that the two samples' responses are similar. While the level of support might change to some extent for individual rights, the overall perceptions of AI and robot rights and the degree to which people favor or disfavor certain rights are similar.  

\section{Discussion on AI and Robot Rights}

This section will review the findings on each of the rights addressed in this research. We discuss the results from three themes that emerge: rights that are most likely observed in \textit{legal}, \textit{societal}, and \textit{private} domains. We only discuss trends that are statistically significant at $P$\textless0.05. We present the standardized effect size statistic $g$ and report the minimum value if there are multiple treatment groups. Please refer to the corresponding figures to examine the average pre-stance (Fig.~\ref{fig:prestance}), the effect size by intervention (Fig.~\ref{fig:intervention}), and the average post-stance (Fig.~\ref{fig:poststance}), respectively.

\subsection{Rights Related to Legal Domains}

The rights to sue and be sued, to hold assets, to enter contracts, and to grant copyright, for instance, are most relevant in the context of legal systems.

\subsubsection{Right to Sue and Be Sued} 

While the participants are initially against allowing AI and robots to sue and be sued (\pre=-0.83), interventions could modify people's perceptions ($g$=0.209, \sch=0.283, \post=-0.548). Artificial persons taking part in lawsuits are common, and after providing more information---particularly the non-human design that explicitly mentioned corporations could sue and be sued---contributed to a more supportive position towards this right. 
The stance was dependent on how the right was phrased. When the right was introduced indirectly (e.g., ``Someone should be able to sue and be sued by robots for injuries and damages''), the participants demonstrated a higher level of support ($g_{pre}$=0.197, $g_{post}$=0.166) than when it was articulated explicitly. 
The respondents were marginally more supportive of granting this right to AI compared to robots ($g_{pre}$=0.183, $g_{post}$=0.117). The same positive effect is found on the level of autonomy; fully autonomous AI and robots are seen as more worthy of the right than when autonomy was not emphasized ($g_{pre}$=0.135, $g_{post}$=0.140).

\subsubsection{Right to Hold Assets} 

The participants are initially opposed to granting assets to AI and robots (\pre=-0.980); our interventions are moderately successful in modifying people's perception ($g$=0.252, \sch=0.322, \post=-0.658). The initial support towards the right to hold assets was also dependent on how the right was phrased. The participants reported a more supportive pre-stance ($g_{pre}$\textgreater0.157) whenever the right was introduced indirectly (e.g., ``assets can be possessed by robots'') rather than directly or explicitly.  

\subsubsection{Right to Enter Contracts} 

Participants are mostly opposed to granting this right to electronic agents (\pre=-0.863). Similarly to the case of the rights previously discussed, people's perception was somewhat influenced by the interventions ($g$=0.222, \sch=0.292, \post=-0.566). The non-human intervention achieved notable success in terms of stance-change ($g$=0.442, \sch=0.582); we posit that this effect was partially caused by the fact that the right to enter contracts was explicitly introduced as a right of corporations in the proposed non-human intervention. Similarly to the other rights associated with the relation of an entity with the legal system, the right to enter contracts was more supported by the participants ($g_{pre}$\textgreater0.462, $g_{post}$\textgreater0.164) when articulated indirectly (e.g., ``Valid contracts can have robots as parties'').

\subsubsection{Right Granted Under Copyright Law} 

The participants are not extremely contrary to granting the rights of an original work to AI and robots (\pre=-0.392). Interventions were also marginally successful in modifying people's perception of the issue ($g$=0.099, \sch=0.139, \post=-0.253). In particular, the non-human intervention was extremely effective, almost resulting in a more supportive attitude ($\max$ \post=-0.026). The form in which the right was introduced to the participants again influenced their perception. Unlike the other previously discussed rights, explicit mentions (e.g., ``Robots should have the rights granted under copyright law'') resulted in a more positive pre-stance towards the right ($g_{pre}$\textgreater0.184). The level of autonomy of the electronic agent again influenced the participants' post-stance; emphasizing full autonomy led to a more supportive position ($g_{post}$=0.154).

\subsection{Rights Related to Societal Domains} 

Some of the proposed rights are most relevant in the context of social interactions, such as the rights to freedom of speech, to a nationality, to choose occupation freely, and to remuneration.

\subsubsection{Right to Freedom of Speech} 

The participants have shown not to be extremely contrary to the idea of granting freedom of speech to AI and robots (\pre=-0.391). The effect of stance-change was similar across interventions, and the maximum post-stance was close to zero ($g$=0.152, \sch=0.212, \post=-0.180; $\max$ \post=-0.092), indicating that people are marginally negative or nearly neutral towards the right to freedom of speech to AI and robots upon an intervention. However, we note there might be sampling bias in that all participants are residents of the US, a country where such a right is widely supported. People's pre-stance was influenced by how the right was articulated. Explicit phrasing (e.g., ``Robot should have the right to freedom of speech'') led to a lower degree of support ($g_{pre}$\textless-0.184). Post-stance, on the other hand, was influenced by how the agent was introduced in terms of autonomy; explicitly introducing full autonomy led to a higher post-stance upon intervention ($g_{post}$=0.130).

\subsubsection{Right to a Nationality}
The participants are initially contrary to the idea of granting nationality to electronic agents (\pre=-0.807). Interventions could incur a close to moderate level of stance change ($g$=0.223, \sch=0.294, \post=-0.513). Pre-stance towards this right depended on how it was phrased; explicitly mentioning it (e.g., ``Robots should have the right to a nationality'') led to the smallest level of support ($g_{pre}$\textless-0.065) compared to other kinds of phrasing.

\subsubsection{Right to Choose Occupation Freely}
The respondents do not think that AI and robots should choose their occupations (\pre=-0.739). Interventions were also not highly effective in changing people's perception of the issue ($g$=0.150, \sch=0.197, \post=-0.543). We posit that participants imagine AI and robots as specialized workers, an image not that far from the current situation. Electronic agents are often built for a designated set of functions, and Artificial General Intelligence (AGI) is nowhere close to being developed~\cite{bostrom2017superintelligence, aginowhereclose}. Once more, the phrasing type impacted the responses; an indirect introduction (e.g., ``No one can coerce AI programs to complete any work'') caused a higher level of support for both pre-stance and post-stance ($g_{pre}$\textgreater0.360, $g_{post}$\textgreater0.105).  

\subsubsection{Right to Remuneration}
The participants are largely contrary to the idea of paying robots and AI for their work (\pre=-0.801). Interventions were however somewhat successful and led to positive stance-changes ($g$=0.222, \sch=0.293, \post=-0.508).  Phrasing only impacted responses related to pre-stance towards the right; explicit phrasing (e.g., ``Robots should have the right to receive payment for their work'') was the most supported ($g_{pre}$\textgreater0.075). We hypothesize that compensating an electronic agent for its work is unreasonable because of the initial investment needed for the AI or robot when acquiring an electronic entity. They are a property that can be bought, and participants do not wish to bear any extra cost.

\subsection{Rights Related to Private Domains}
Finally, the right to privacy, to life, and against cruel treatment and punishment are most relevantly associated with the \textit{``private''} boundary of an electronic agent.

\subsubsection{Right to Privacy}
 
The participants are mostly contrary to the idea of granting the right to privacy to AI and robots (\pre=-0.577). The proposed intervention also reported a marginal effect ($g$=0.124, \sch=0.169, \post=-0.408), showing that people do not see the privacy of AI and robots as necessary. This is in line with the current GDPR that requires complete transparency of systems (and hence no privacy for AI and robots). 
Again, the participants' pre-stance was subject to how the right was phrased; explicit phrasing (e.g., ``Robots should have the right to privacy'') led to a lesser degree of support towards the right ($g_{pre}$\textless-0.182).

\subsubsection{Right to Life}
 
The participants are strongly against preserving the ``life'' of a robot or AI (\pre=-0.964), yet alone interventions had a close to moderate effect ($g$=0.235, \sch=0.296, \post=-0.668). Pre-stance towards the right to life of AI and robots showed a significantly lower average value ($g_{pre}$\textless-0.213) if the right was indirectly introduced (e.g., ``No one should be able to turn off or kill robots'') rather than directly or explicitly. 
Additionally, explicitly introducing full autonomy led to significantly higher post-stance values towards electronic agent's right to life ($g_{post}$=0.160).

\subsubsection{Right Against Cruel Punishment and Treatment}

Pre-stance results indicate that the participants are initially supportive of protecting automated agents against cruelty (\pre=0.171). None of our proposed interventions were significant in changing people's perception regarding this right ($g$=0.028, \sch=0.038, \post=0.209), likely because there is little room to shift even more positively. This finding resonates with anecdotal evidence presented by Darling~\cite{darling2016extending, darling2015s}, which found humans feeling pity for robots or refusing to ``hurt'' them. Darling proposed to create legal protections against the harsh treatment of robots, not for the sake of robotic agents but as a form of discouraging cruel behavior to be transferred to human interactions. Whenever the right against cruel treatment was introduced explicitly (e.g., ``Robots should have the right against cruel punishment or treatment''), however, the participants reported a more substantial degree of rejection to the proposal ($g_{pre}$\textless-0.389, $g_{post}$\textless-0.177). We thus posit that the participants agree that cruelty against AI and robots should not be accepted; granting a right to protect them, however, is not seen as a reasonable alternative.

\section{Implications}

The possibility of granting rights to AI and robots has various implications for the development of these systems. Creating a product---as robots and AI are currently treated---comes with specific responsibilities that might lead to some ethical and legal issues, such as product liability impositions. However, creating a "person" with its rights and responsibilities brings new responsibilities that might not be initially expected by developers, users, and manufacturers. Such responsibilities need to be discussed proactively than reactively. Below we discuss implications for the three rights that received the highest average post-stance.
 
First is the right against cruel punishment and treatment. Online users gave positive support during both pre-stance and post-stance surveys on this right, even though electronic agents cannot feel pain.\footnote{While pain needs a clear definition in some fields, we use its agreed-upon definition based on human or animal sentience.} This result is in line with the widespread sentiment toward the viral video inspired by the Boston Dynamics' robot that was kicked over by humans.\footnote{\url{https://youtu.be/dKjCWfuvYxQ}} Thousands of online users called this an "abusive" action, although the action was theoretically meant for testing~\cite{robotsbeaten}. As a direct implication, future AI and robot designers and manufacturers might need to consider this right during the development process. Systems could be in place to report cruel behavior to authorities should robots or AI were to be granted such right. Another possibility would be alarming cruel treatment against AI and robots, such as demonstrating pain and agony regardless of the lack of such qualities in these systems.

Second is the right to copyright for original work, which received a nearly neutral decision in the post-stance results. This particular right can be framed in the form of electronic agents with generative models, such as AI writers and composers, to generate creative work in many areas. Questions regarding who should own such original work created by AI or robots would be highly contested~\cite{copyrightgan}, since a system's creation would be its own and not belong to the manufacturer, developer, or owner, possibly leading to various copyright infringements lawsuits. This possibility conflicts with the current laws, as exemplified by the rejection of two AI-invented patents by the UK and EU patent offices under the justification that the inventor was not a human~\cite{iprightsrejection}. Some researchers are working on machine learning watermarks~\cite{wang2019robust, zhong2019robust}, which could also clash with these systems' copyright.

The third is freedom of speech. In the domain of journalism, fake news and misinformation may be distributed more broadly or may even be initiated by AI writers, as these systems take even more crucial roles in journalism~\cite{carlson2015robotic}. Regulation of AI reporters, therefore, might clash with their possible freedom of speech. Additionally, developers might need to consider a generative model's freedom of speech when deciding whether it should be deployed. Take, for instance, Open AI's decision not to instantaneously release its GPT-2 model~\cite{radford2019language} due to concerns regarding its possible uses~\cite{openaiwriter}. Similar decisions in the future might conflict with electronic agents' freedom of speech, had this right been granted to future electronic agents.

As discussed above, AI and robot rights would raise various legal and moral questions that have yet to be answered. However, these rights could also change how people perceive, and thus interact, not only with these systems but with society as a whole. Rights are attached to the imposition of social obligations and duties~\cite{solaiman2017legal}. Therefore, granting an individual moral and legal status to electronic agents should change how society is organized. The abolishment of slavery, namely the moral consideration of previously neglected entities, heavily modified the social fabric. While Bryson addresses some negative societal consequences of treating these AI and robots as more than mere tools~\cite{bryson2010robots}, it is expected to change even more if rights are to be granted. With this work, we call for CSCW and HCI researchers to address these questions. Human-machine and human-human interactions are expected to drastically change as these systems are deployed, and these changes are to be more extreme if we grant certain moral and legal considerations to AI and robots.

Our results on the public perception of AI and robot rights can also offer useful guidelines for engineers, entrepreneurs, and policymakers who attempt to introduce automated agents to collaborate with humans in various workspaces. It is expected that an increasing number of robots and AI systems will be employed in situations that require the cooperation between humans and non-human agents in physical proximity or remotely~\cite{levy2012new, vertesi2015seeing}. The human workers will have to be informed verbally or textually about the nature, capability, and limitation of the non-human agents whenever and wherever this happens. To ensure the effective and safe incorporation of these non-human automated agents, introducing them into a working relationship with humans will have to be codified in the form of manuals, guidebooks, or training materials. Through these linguistic devices, human workers perceive and characterize automated agents as their co-workers, subordinates, superiors, friends, or enemies.

The most relevant finding from our survey in this regard is that the styles, phrases, and metaphors used in these materials impact how the human workers understand the robotic workers. Our intervention examples suggest that certain styles of describing AI and robot agents can even lead human workers to modify their initial responses to these agents, granting them more or fewer rights and duties. According to our study, it will be necessary to craft the words carefully for explaining the automated agents to the human co-workers so that they can manage their expectations of those agents' capability or responsibility vis-à-vis their own. This will be even more important for AI and robots employed for safety-critical missions in extreme environments such as outer space or the war, where the issues of accountability and punishment frequently emerge ~\cite{carpenter2016culture,mindell2015our}. Above all, our analysis of the public perception of AI and robots shows that the co-work or cooperation between humans and autonomous electronic agents is fundamentally social~\cite{hutchins1995cognition}. It means that humans can work with automated agents properly by making sense of them---their identities, rights, and responsibilities---through language, appearance, or gesture. As more automated computing agents move into our workplaces, social and linguistic understanding will become a significant concern for technology policies in the future.

\section{Future Work}

This work aimed to be a starting point of obtaining the general public's perception of AI and robot rights. We addressed 11 rights for future AI and robots based on a rich set of literature. Many rights granted to both natural and non-natural entities were not covered in this research and must be addressed in the future. Our sample of respondents were US residents, and their attitudes towards AI and robot rights might not be generalizable to the world population. Studies have indicated how public perception of robots is dependent on respondents' cultural background~\cite{nomura2015differences, li2010cross, syrdal2009negative}, and future work should address how culture can affect attitude towards AI and robots rights.

Future research can also delve deeper into the various forms of metaphors, writing styles, and persuasion techniques. For instance, we have not addressed an "indirect and explicit" phrasing style (e.g., "The right to hold assets should be granted to completely autonomous robots."). As discussed in Section 3.2, the premise of this work, namely the concept of legal personhood, is elastic and inclusive of various entities. Therefore, most of the interventions were somewhat one-sided for granting rights to neglected and non-natural entities in society. Understanding how people modify their stance towards the proposal given an opposing idea (e.g., discussing threats arising from granting rights to AI and robots) will be a much needed future work for the continued and holistic crafting of regulations.

Finally, our work suggests that society might not oppose, and even support, some rights to electronic agents. In future work, we aim to understand better what specific aspects of these rights could be granted to these entities and how they could be implemented. For instance, should all robots be granted certain protections from cruelty or only those who are anthropomorphized by society? How can policymakers preserve possible freedom of speech of AI journalists while protecting society from harmful and fake journalism? Granting rights to a specific entity does not come by itself but accompany responsibility and duties to these same entities. Future studies should also address how the general public assigns responsibility to electronic agents for their actions, both in cases of complete autonomy and shared-responsibility ~\cite{awad2019drivers,lima2020punishing}. 

\section{Concluding Remarks}

This study collected online users' perception of 11 possible rights for highly advanced and autonomous AI and robots of the future. We tested flexibility in people's responses by exposing them to four kinds of carefully designed interventions. Our study shows that online users, at first, disfavor rights to electronic agents. Those rights most relevant in the context of legal systems and life --- i.e., the right to sue and be sued, the right to hold assets, the right to enter contracts, and the right to life --- faced the highest rejection amongst all. The only exception was the right against cruel treatment and punishment. 
 
The experiment also showed whether and how online users' perceptions can be modified. Respondents took more positive views after being exposed to interventions that corrected misinformation on legal personhood. This change was statistically significant. Listing examples of rights-bearing non-human entities (i.e., non-human design) and explicitly explaining that humanity is not a requirement to hold rights (i.e., descriptive design) were the two most effective. On the other hand, discussing the lack of agreement among scholars on the requirements for legal personhood (i.e., normative design) and listing examples of human entities who were socially neglected in the past (i.e., human design) did not, in many cases, result in a significant stance-change. 

Phrasing style was an essential factor that affected how people responded to our survey. For most of the proposed rights, indirectly introducing the proposal elicited more support. We posit that an allusive discussion style may create a sense that, rather than a right is granted to an electronic entity, a new duty is created for existing legal persons. In some scenarios, however, explicit phrasing elicited more support. We hypothesize that the concept of ``rights'' depends on its context and how it is posed. This finding is in accord with the argument that the regulation of AI will depend on which metaphors are used during the process~\cite{richards2013should}. Likewise, the metaphors on how the rights were phrased affected people's judgment of AI and robot rights. Explicitly reminding people that the target electronic agents are completely autonomous was also critical in bringing more support; the stance towards most rights became more positive under this writing style.

Finally, we discuss a possible subject-expectancy effect in the survey results, a form of bias that might arise due to the reactivity of research subjects in the experiment. Since the participants were presented with similar survey questions before and after the intervention, they may have been motivated to change their responses regardless of the intervention message. However, we consider this bias would be minimal for two reasons. First, the intervention was not instantaneous and took on average 97.5 seconds (see Appendix B). This indicates that participants spent enough time reading and digesting the content. Second, the stance-change was not random but consistently positive after the intervention. Had any subject-expectancy effect been in play, the stance-change direction would not be consistent across all rights.

Our research does not, by any means, propose AI and robots be treated as humans, or as other non-natural legal persons. Electronic agents are a new type of entity that must be regulated from the start while taking past experiences with diverse bodies and technologies. Our survey results resonate with the argument that regulations should not necessarily consider AI and robots as ``humans,'' but one must acknowledge that people might treat them as so~\cite{richards2013should}. 

Granting legal personhood to electronic agents is a contested possibility, and our results indicate that people are somewhat flexible and supportive of specific rights for AI and robots. When considered along with the previous progression of rights for animals and corporations (i.e., non-human entities), our results suggest that social norms could take one day shift in favor of granting fully autonomous AI and robots some rights (at least one or several) discussed in this paper. Furthermore, given the controversial aspect of the proposal and historical precedents (e.g., rights of women and slaves), which have shown that granting rights to previously neglected entities is a slow process, our finding supports the argument that public opinion might change in the future. Considering that our intervention was simple, real-life exposures to various other interventions may lead to a more considerable stance change.

In conclusion, debunking common misconceptions regarding rights and duties to non-natural entities have shown to be successful in modifying people's stance towards the possibility of granting rights to AI and robots. Our results also indicate that some rights to electronic agents might be supported by the general public soon such as, for instance, the rights against cruelty and copyright. We hope this research promotes a more inclusive discussion of a prominent, yet controversial issue involving the rapid deployment of AI and robots in society. Granting rights to non-humans (and even some groups of humans) has always faced prejudice and rejections at first thought, but doing so has become an essential social and political aspect of contemporary society \cite{teubner2006rights}.

\section*{Acknowledgments}

G. Lima, C. Kim, S. Ryu and M. Cha were supported by the Institute for Basic Science (IBS-R029-C2) and Basic Science Research Program through the National Research Foundation Korea of Korea (No. NRF-2017R1E1A1A01076400). C. Jeon was supported by the framework of international cooperation program managed by National Research Foundation of Korea (NRF-2018K1A3A7A03089893).


\bibliographystyle{ACM-Reference-Format}
\bibliography{sample-base}

\newpage 
\appendix

\section{Mean Stance-Change}

The figure below shows the average stance-change (i.e., \sch) in the x-axes across all 11 rights, and the y-axes indicate the intervention type. These results are discussed in Sections 4.2 and 6.

\begin{figure*}[h]
    \centering
    \includegraphics[scale=0.53]{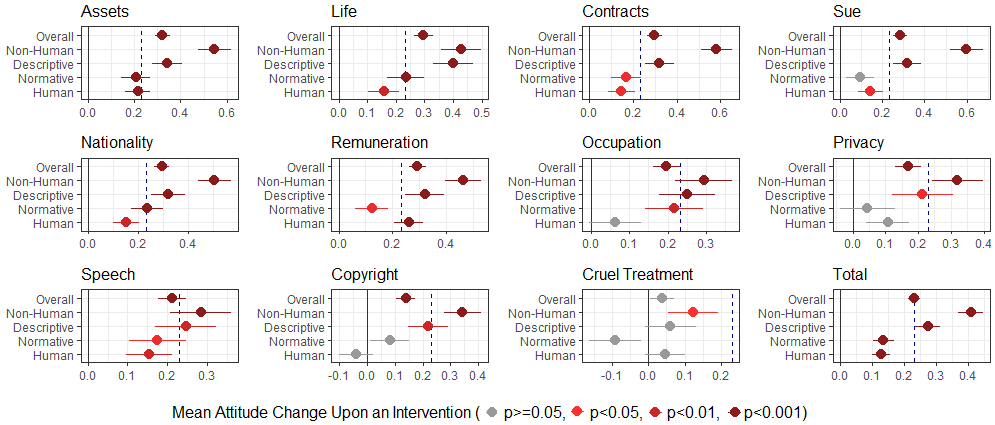}
    \caption{Mean stance-change (\sch) upon each intervention, where colors indicate level of significance. The blue dashed line represents the mean stance change across all interventions. The gray line indicates zero, i.e. null stance-change. 
    }
    \label{fig:interventionappendix}
\end{figure*}

\section{Time Spent on Interventions}

The figure below shows the cumulative distribution of time spent on interventions from Section 3.4. The average time spent on the intervention is 97.5 seconds, and the minimum time spent on the intervention is 27 seconds. 

\begin{figure}[htpb!]
    \centering
    \includegraphics[width=0.4\textwidth]{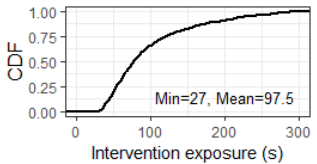}
    \caption{Cumulative distribution of time spent reading interventions for the MTurk sample ($N$=1270).}
    \label{fig:interdist}
\end{figure}

\newpage
\section{Experimental Design}

Our experimental designs, both declarative statements and intervention cards, are available at \url{https://github.com/dscig/AIRights_PublicPerception}. The figure below exemplifies how the experiment was presented to participants in Amazon Mechanical Turk (a) and Qualtrics (b). These two designs contain the same content, but have different look and feel. Due to these differences, we do not attempt to directly compare the stance values across the two platforms in Section~5. 

\begin{figure}[htpb!]
     \centering
     \begin{subfigure}{\textwidth}
         \centering
         \includegraphics[width=0.7\textwidth]{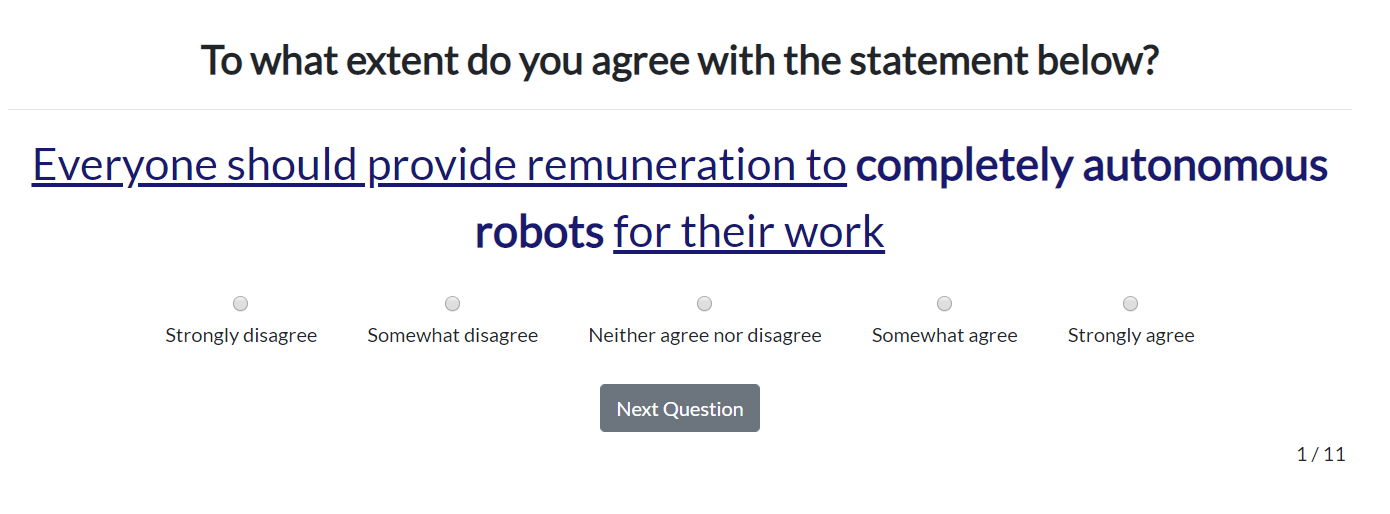}
         \caption{Survey design screenshot on Mturk.}
     \end{subfigure}
     \par\bigskip
     \begin{subfigure}{\textwidth}
         \centering
         \includegraphics[width=0.7\textwidth]{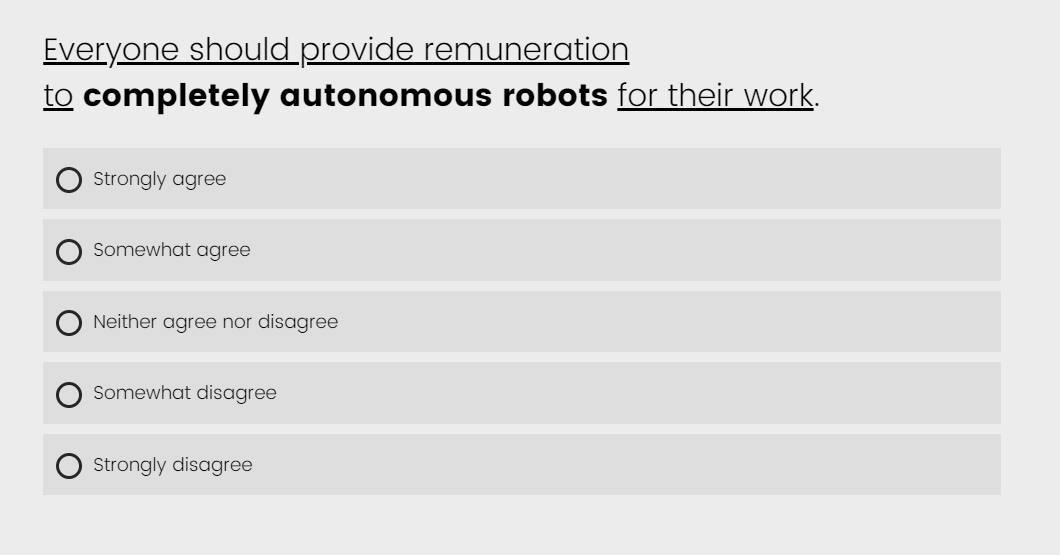}
         \caption{Survey design screenshot on Qualtrics.}
     \end{subfigure}
     \caption{Screen captures of a statement presented to participants recruited through (a) the Amazon Mechanical Turk (MTurk) and (b) Qualtrics.}
     \label{fig:screenshots}
\end{figure}








\end{document}